\documentclass[10pt,a4paper]{article}

\usepackage{times}

\usepackage{epsfig,wrapfig}
\usepackage{epstopdf} 

\usepackage{sidecap}
\usepackage{graphicx}
\usepackage{color}

\usepackage{fancyhdr}

\usepackage[lmargin=2.5cm, rmargin=2.5cm,tmargin=3.50cm,bmargin=3.50cm]{geometry}
\usepackage{indentfirst}

\usepackage{amsmath}
\usepackage{amssymb}
\usepackage{bbold}

\usepackage{titlesec}
\titleformat{\section}[hang]
  {\centering}{\thesection}{1ex}{\normalsize \textsc}
\titleformat{\subsection}[hang]
  {}{\thesubsection}{1ex}{\normalsize \textit}

\renewcommand{\thesection}{ \normalsize \textnormal{\Roman{section}.}}
\renewcommand{\thesubsection}{\normalsize \textnormal{\textsc{\textit{\Alph{subsection}.}}}}

\def\e{\begin{equation}}
\def\f{\end{equation}}
\def\_#1{{\bf #1}}

\def\.{\cdot}

\begin{document}

\title{\large \textbf{Spectral Design of Active Mechanical and Electrical Metamaterials}}
%
\def\affil#1{\begin{itemize} \item[] #1 \end{itemize}}
\author{\normalsize \bfseries H. Ronellenfitsch$^1$ and \underline{J. Dunkel}$^1$}
%
\date{}
\maketitle
\thispagestyle{fancy} 
\vspace{-6ex}
\affil{\begin{center}\normalsize $^1$Massachusetts Institute of Technology, Department of Mathematics, 77 Massachusetts Ave, Cambridge, MA 02139, U.S.A.\\
dunkel@mit.edu
 \end{center}}

\begin{abstract}
\noindent \normalsize
\textbf{\textit{Abstract} \ \ -- \ \
Active matter is ubiquitous in biology and becomes increasingly more important in materials science.
While numerous active systems have been investigated in detail both experimentally and theoretically, general design principles
for functional active materials are still lacking. Building on a recently developed linear response optimization (LRO) framework, we here demonstrate that the spectra of
 nonlinear active mechanical and electric circuits can be designed similarly to those of linear passive networks.}
\end{abstract}

\section{Introduction}
Active networks model nonequilibrium systems across a wide
range of scales, from the cytoskeleton~\cite{Broedersz2014} to traffic
flow~\cite{Coclite2005}. Metamaterials have similarly
broad applications in engineering, physics and art, ranging from
acoustics~\cite{Deymier2013,Cummer2016} to sonic sculptures~\cite{Martinez-Sala1995},
and antennas~\cite{Soric2014,Minatti2015}.
While substantial progress has been made in the design of acoustic materials~\cite{Bendsoe2003},
the functional optimization of active metamaterials has remained less explored so far.
Here, we bridge this gap by applying inverse LRO design techniques~\cite{Ronellenfitsch2018} for discrete
metamaterials to two generic
classes of active networks.
Specifically, we consider mechanical and electric circuits, consisting of interconnected active nodes driven by
generic nonlinearities, and show that inverse bandgap-design of the underlying linear network structure
can suppress active nonlinear oscillations at prescribed frequencies.

\begin{figure}[h!]
\centering
\epsfig{file=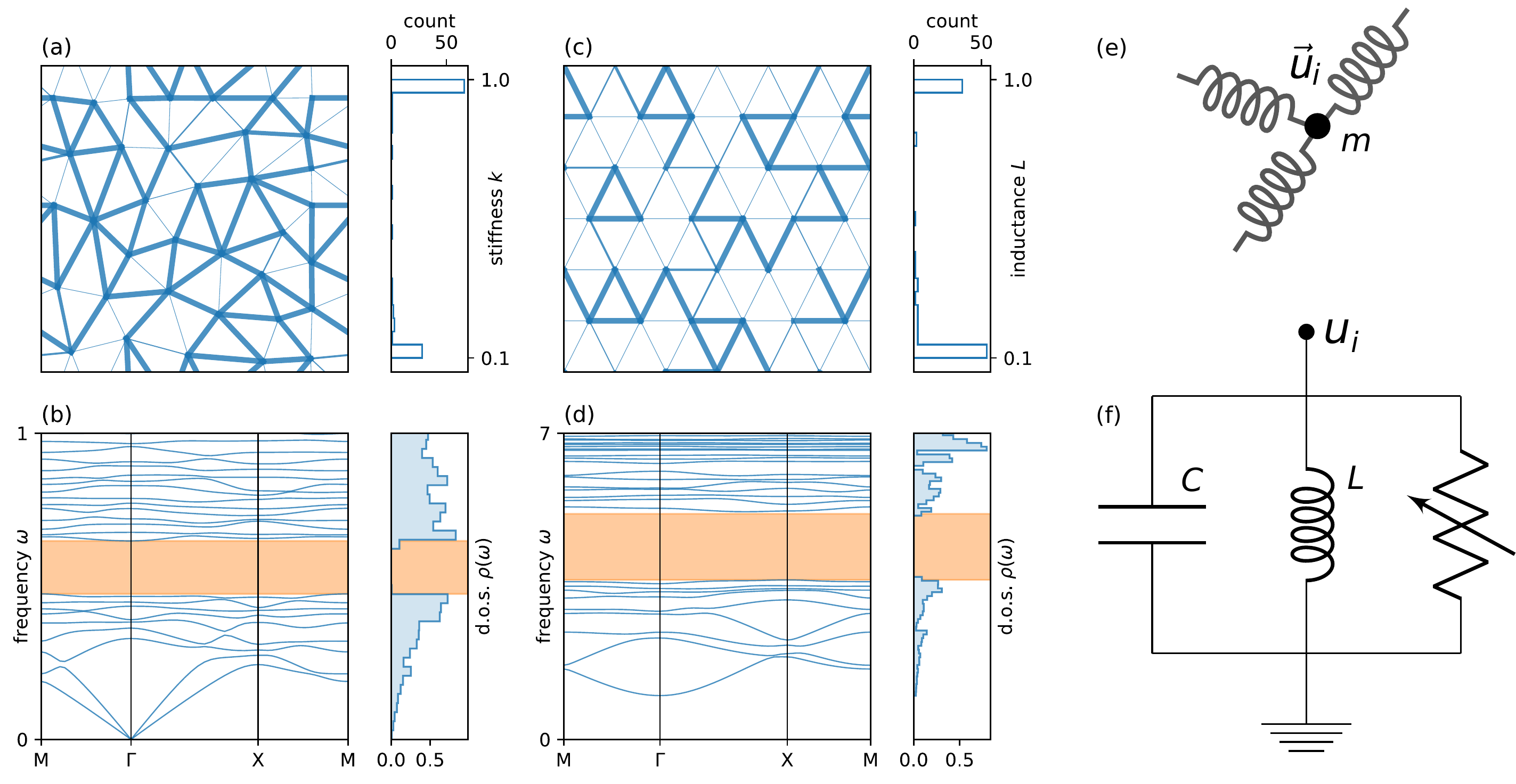, height=0.3\textheight}
\caption{\small{Response-optimized mechanical and electrical networks.
(a) Unit cell of a mechanical network  on a randomized
topology. Line thickness corresponds to spring stiffness.
(b) Phononic band structure and density of states (d.o.s.) of the network
from (a); the tuned band gap is marked in orange.
(c) Unit cell of an electric LC circuit on a triangular grid.
Line thickness corresponds to connecting inductance.
(d) Band structure and d.o.s. of the network shown in (c);
the tuned band gap is marked in orange.
(e) Mechanical mass point with displacement $\mathbf{u}_i$ and mass $m$
connected by three springs to the rest of the network.
(f) Electrical oscillator node with voltage $u_i$, capacitance $C$,
inductance $L$, and a nonlinear resistance.}} \label{fg:fig1}
\end{figure}

\section{Model and Results}
As models directly amenable to inverse design~\cite{Ronellenfitsch2018}, we
consider linear second-order dynamical systems,
\begin{align}
    \ddot{\mathbf{u}} + D \mathbf{u} = 0,
    \label{eq:network}
\end{align}
where the vector $\mathbf{u}$ contains the degrees of freedom and $D$ is a
positive semi-definite dynamical matrix. Specifically, we focus on
two-dimensional linearized mechanical balls-and-springs networks~\cite{Ronellenfitsch2018} and
electrical LC oscillators
connected by inductors~\cite{Kotwal2019}. In the mechanical case, the dynamical matrix $D_\text{mech} = m^{-1}QKQ^\top$,
where $Q$ is the equilibrium matrix~\cite{Lubensky2015}, $K$ is a diagonal matrix of spring
stiffnesses and $m$ is the particle mass
(Fig.~\ref{fg:fig1}e).
In the electric circuit case, the dynamical matrix is
$D_\text{elec} = {C}^{-1}(L^{-1}\mathbb{1} +  EWE^\top)$, where $L$ is the inductance of each oscillator,
$C$ is its capacitance, $E$ is the network's oriented incidence matrix, and the diagonal
matrix $W$ contains the inverse inductances connecting the oscillators (Fig.~\ref{fg:fig1}f).

Both  circuit  systems naturally support wave propagation through
phonons and electrical oscillations, respectively. In large periodic crystals, the possible modes
of oscillation are described by a band structure.
If this band structure is tuned, waves and oscillations
can be controlled. Applying the LRO method described in Ref.~\cite{Ronellenfitsch2018},
we designed mechanical and electric circuits by tuning spring stiffnesses $K$ and
inverse inductances $W$, respectively.
The underlying unit cells were designed to exhibit large bandgaps around a desired frequency
of oscillation (Fig.~\ref{fg:fig1}a--d) by minimizing the averaged linear response
to forcing at frequency $\omega$,
$R(\omega) = \operatorname{tr}\left(G(\omega) G(\omega)^H\right)$,
where $\operatorname{tr}(\cdot)$ is the matrix trace and $G(\omega) = (-\omega^2\mathbb{1} + D)^{-1}$
is the linear response matrix.
We now show that this network design directly carries over
from linear passive networks to non-linear active networks, where each node is endowed
with the ability to convert some fuel available in the environment into motion.

\begin{figure}[b!]
\centering
\epsfig{file=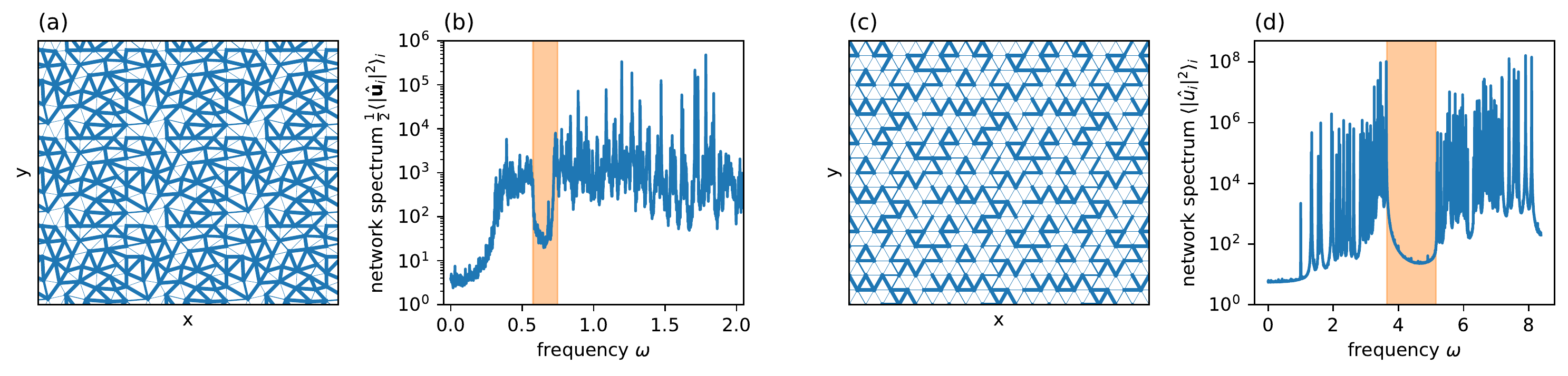, width=.825\textwidth}
\caption{\small{Nonlinear active circuits based on the tuned unit cells from Fig.~\ref{fg:fig1}.
(a,c) Mechanical and electrical finite-period networks consisting of
$3 \times 3$ unit cells.
(b,d) Temporal Fourier spectrum averaged over all
active mechanical/electrical oscillators in the network. The location of the bandgaps
in corresponding linear networks is marked in orange. Activity
parameters for the mechanical network in (a,b)
were $\kappa=0.1, m=1, \gamma=0.1, \gamma_f=0.2, v=1$. For the electrical network in (c,d) we chose
$\rho=0.01, \varepsilon=0.02$.}}
\label{fg:fig2}
\end{figure}

A generic model for active mechanics is obtained by adding a non-linear
friction force to each node~\cite{Woodhouse2018,Romanczuk2012},
\begin{align}
    \mathbf{F}_i = -\gamma\dot{\mathbf{u}}_i + \gamma_f (1 -
    |\dot{\mathbf{u}}_i |^2/v^2) \dot{\mathbf{u}}_i .
    \label{eq:Rayleigh}
\end{align}
Here, $\gamma$ is a regular friction coefficient, the two-dimensional vectors
$\mathbf{u}_i$ and $\dot{\mathbf{u}}_i$ are the displacement and velocity of the
$i$th node, $\gamma_f$ is a measure of the nonlinearity, and $v$ is the
particle's preferred velocity.
A free particle following Eq.~\eqref{eq:Rayleigh} will come to rest
if $\gamma > \gamma_f$ or approach the speed $|\dot{\mathbf{u}}_i| = v$
in some direction if $\gamma < \gamma_f$.

Although it is possible to obtain band structures of non-linear systems
in certain cases~\cite{Narisetti2011}, we use here an alternative approach that allows us to
study the allowed modes of oscillations in a direct numerical manner.
To this end, we simulated the active force Eq.~\eqref{eq:Rayleigh} with the designed network
dynamics Eq.~\eqref{eq:network} from Fig.~\ref{fg:fig1}(a) for
times $t=[0,10^4]$ from Gaussian random initial conditions and obtained
time series $\mathbf{u}_i(t)$. To prevent the individual oscillators from
performing collective translational motion, we fixed them in place using
uniform springs of strength~$\kappa$ by setting $D_\text{mech} \to \kappa m^{-1}\mathbb{1}
+ D_\text{mech}$.
To extract the active vibrational spectrum we ignored initial transients
and computed the discrete Fourier
transform $\hat{\mathbf{u}}_i(\omega)$ of the time series. We then averaged the
Fourier spectrum over all nodes of the network to obtain the active
network spectrum $\frac{1}{2}\langle|\hat{\mathbf{u}}_i|^2\rangle_i$.
The nonlinear active system shows a pronounced dip
in its spectrum near the linearly designed bandgap (Fig.~\ref{fg:fig2}b).
This suggests that the LRO-designed network structure is able to
suppress nonlinear active motion at its gapped frequencies.

In a similar manner, we now consider active
electrical circuits coupled to the designed network from Fig.~\ref{fg:fig1}(c).
Following Ref.~\cite{Kotwal2019}, we study the canonical nonlinear
van der Pol relaxation oscillator~\cite{VanderPol1926} in the regime of weak
nonlinearity. Coupling the nonlinear van der Pol resistance with the network
dynamics Eq.~\eqref{eq:network}, we obtain the equations of motion for the dimensionless nodal voltages $u_i$,
\begin{align}
    \ddot{u}_i + \rho \dot{u}_i - \varepsilon (1 - u_i^2) \dot{u}_i
        + \sum_j D_{ij} u_j = 0,
\end{align}
where $\varepsilon$ is a measure of the strength of the nonlinearity and we introduced
an additional resistance $\rho$ to make formal contact with the active mechanical force
Eq.~\eqref{eq:Rayleigh}. If $\rho < \varepsilon$, the circuit becomes active
and produces self-sustained oscillations.
Repeating the numerical procedure from before, we obtained
spectra for the active electrical network and again find that the linearly designed
bandgap carries over to the active system (Fig.~\ref{fg:fig2}d). This shows that the passive linear network
structure can control and suppress active nonlinear dynamics in the case
of electric circuits as well. Our numerical experiments suggest, however, that in the van der Pol network, this
is only possible for weak nonlinearity ($\varepsilon \lesssim 0.05$) whereas the mechanical
network can control much larger nonlinear active forces.

\section{Conclusions}
We have demonstrated that both linear mechanical and electrical networks can be inversely
designed using LRO optimization methods~\cite{Ronellenfitsch2018},  and that designed spectral gaps and
accompanying suppression of motion at certain frequencies carry over to
nonlinear active networks. The networks considered above are prototypical
examples of active metamaterials, and we expect that inverse design of such
materials will find fruitful applications in engineering, physics, and biology.



{\small
\bibliographystyle{unsrt}
\bibliography{references}{}

}

\end{document}